\documentclass[aps,pra,twocolumn,preprintnumbers,amssymb,amsfonts,amsmath,superscriptaddress,showpacs,hyperref]{revtex4-1}

\pdfoutput=1
\usepackage{amssymb}
\usepackage{bm}
\usepackage{cancel}
\usepackage{color,graphicx}
\usepackage{amsmath}
\usepackage{relsize}
\usepackage{amsbsy}
\usepackage{amsthm}
\usepackage{bbm}
\usepackage{epsfig}
\usepackage{float}
\usepackage{graphicx}
\usepackage{dcolumn}
\usepackage{bbm}
\usepackage{color,epstopdf}
\usepackage{amscd}
\usepackage{amsfonts}  
\usepackage[]{amsmath}
\usepackage{amssymb}    
\usepackage{mathrsfs}
\usepackage{verbatim}
\usepackage[]{cases}
\usepackage{amsmath}
\usepackage{wasysym}
\usepackage[utf8]{inputenc}
\usepackage[T1]{fontenc}
\usepackage{mathtools}
\usepackage{xcolor}
\usepackage{tabularx}

\usepackage{soul}

\newcommand{\hide}[1]{\relax}

\newcommand{\nocontentsline}[3]{}
\newcommand{\tocless}[2]{\bgroup\let\addcontentsline=\nocontentsline#1{#2}\egroup}

\begin{document}

\title{Optomechanics-based quantum estimation theory for collapse models}

\author{Marta Maria Marchese}
\affiliation{Department of Physics and Astronomy, The University of Sheffield, Hounsfield Road, S3 7RH Sheffield, United Kingdom}
\author{Alessio Belenchia}
\affiliation{Institut für Theoretische Physik, Eberhard-Karls-Universität Tübingen, 72076 Tübingen, German}
\affiliation{School of Mathematics and Physics, Queens University, Belfast BT7 1NN, United Kingdom}
\author{Mauro Paternostro}
\affiliation{Centre for Quantum Materials and Technologies, School of Mathematics and Physics, Queens University, Belfast BT7 1NN, United Kingdom}

\begin{abstract}
We make use of the powerful formalism of quantum parameter estimation to assess the characteristic rates of a Continuous Spontaneous Localisation (CSL) model affecting the motion of a massive mechanical system. We show that a study performed in non-equilibrium conditions unveils the advantages provided by the use of genuinely quantum resources -- such as quantum correlations -- in estimating the CSL-induced diffusion rate. In stationary conditions, instead, the gap between quantum performance and a classical scheme disappears. Our investigation contributes to the ongoign effort aimed at identifying suitable conditions for the experimental assessment of collapse models. 
\end{abstract}

\maketitle
\section{Introduction}

The phenomenology of the quantum-to-classical transition, which is the process that drives an otherwise quantum system towards a fully classical description of its physical configuration, is the object of an extensive body of research. Indeed, whether such transition is due to new fundamental physics or not is a controversial matter~\cite{bell_aspect_2004}.

In particular, it is still under debate if the decoherence of a quantum system that grows in complexity and size can be ascribed to an intrinsic mechanism or only to the unavoidable presence of the surrounding environment~\cite{zurek1991decoherence,schlosshauer2005decoherence}. 

Motivated by the fact that environmental decoherence does not provide a satisfactory solution to the measurement problem, and thus to the quantum-to-classical transition issue, collapse models embody an alternative theoretical framework~\cite{Ghirardi1986,bassi2013models}. By elevating the collapse of the  wavefunction to a universal physical mechanism embedded in a stochastic dynamics, collapse models explain the quantum-to-classical transition in a phenomenological fashion, thus embodying an instance of macrorealistic modifications of quantum mechanics. 

Such modification is achieved through a stochastic Sch\"odinger equation and the introduction of new fundamental parameters. When used to assess the dynamics of microscopic systems, the framework of collapse models recovers standard quantum mechanics. Moving towards larger systems, coherence is rapidly suppressed to prevent large spatial superpositions of macroscopically distinguishable states.

The Continuous Spontaneous Localisation (CSL) is one of the most well-studied collapse models~\cite{ghirardi1990markov,bassi2003dynamical}. It describes the loss of coherence in the position basis by way of an an extra dissipative term entering the master equation of a quantum system.
This means that an open quantum system subjected to the collapse mechanism should experience additional dissipation not ascribable to any of the other environmental noise sources.
Testing this model is of current interest for the exploration of the limits of validity quantum mechanics~\cite{Carlesso2022}. However, the predicted collapse effect for most of the systems currently used in quantum labs is very weak and thus challenging to detect and distinguish from other environmental noise effects. 

Entering the regime  where the collapse mechanism is dominant is a tall order. It requires extremely good isolation from environmental noises and ultra-sensitive devices. On the other hand, studies of statistical inference techniques, such as hypothesis testing and parameter estimation, can be employed to ease the requirements and smooth the path towards experimental tests~\cite{mcmillen2017quantum, marchese2021optomechanical}. 
Indeed, in Ref.~\cite{marchese2021optomechanical} we have shown that a quantum hypothesis testing protocol applied to a mesoscopic optomechanical system -- whose massive mechanical  mode would be subjected to the CSL mechanism, if any -- provides advantages with respect to comparable classical strategies and during the transient dynamics before the onset of a stationary state. In this work, we employ the same optomechanical set-up and look at the problem of parameter estimation rather than noise-source discrimination. 

The remainder of this work is organized as follows. In Section~\ref{sec:theory}, we briefly recall elements of quantum parameter estimation theory relevant to our investigation. In Section~\ref{sec:system} we describe shortly the optomechanical set-up of interest. This is a two-cavity system for which quantum advantages stemming from the application of hypothesis testing techniques to collapse model dynamics have been previously shown~\cite{marchese2021optomechanical}. In Section~\ref{sec:dynamics}, we perform a dynamical analysis showing that an advantage, analogous to the one for the hypothesis testing, emerges during the transient also for quantum parameter estimation. In Sec.~\ref{sec:steadystate} we then analyse the steady-state of our optomechanical set-up and show that classical measurement strategies and input noise outperform the quantum strategies considered. We conclude in Sec.~\ref{sec:conclusions} with a discussion of our findings. 

\section{Parameter Estimation Theory}\label{sec:theory}

In Ref.~\cite{marchese2021optomechanical}, we have considered the advantage that arises in a quantum hypothesis testing scenario aiming to test the presence of a collapse mechanism. While a hypothesis testing protocol allows us to determine, up to a certain confidence level, whether something akin to a collapse mechanism is acting upon a system~\cite{schrinski2020quantum}, one can further wonder, how well the collapse parameters can be estimated in principle and which measurement strategies offer the best chances.
Addressing these issues involves the use of quantum estimation theory tools.  

In (local) quantum estimation theory~\cite{giovannetti2006quantum,giovannetti2011advances}, the quantum Cramer-R\'ao bound~\cite{helstrom1969quantum} defines the ultimate precision limit for the estimation of a parameter encoded in the state of the system. Indeed, in general, the parameter of interest ($\Lambda$) does not correspond to a directly measurable observable of the system and its estimation has to be done indirectly, via the measurement of an observable of the parameter-dependent state $\rho(\Lambda)$ of the system. In classical estimation theory, the Fisher information $\mathcal{I}_C(\Lambda)$ provides a lower bound to the mean square error~\footnote{In the following we will consider unbiased estimators for which the mean square error coincides with the variance~\cite{paris2009quantum}} of any estimator of the parameter $\Lambda$. This is known as the (classical) Cramer-R\'ao bound and reads
\begin{equation}
    V(\Lambda)\geq \dfrac{1}{n \mathcal{I}_C(\Lambda)},
\end{equation}
where $n$ is the number of measurements.
The Fisher information is defined as 
\begin{equation}
\begin{split}
    \mathcal{I}_C(\Lambda)=&\int dx\; p(x|\Lambda)\left(\dfrac{\partial \text{ln}p(x|\Lambda)}{\partial\Lambda}\right)^2,
    \label{eq:classicalfisher}
\end{split}
\end{equation}
where $p(x|\Lambda)$ is the conditional probability of obtaining the outcome $x$ when the parameter has value $\Lambda$. It is important to note that this quantity, and thus the classical Cramer-R\'ao bound, depends on the measurement strategy that is adopted to extract information from the state of the system. This is encoded in the conditional probabilities that can be recast in the form of the Born rule $p(x|\Lambda)={\rm Tr}[\Pi_x\rho_\Lambda]$ with $\{\Pi_x\}$ defining the POVM corresponding to the measurement strategy.

The ultimate bound to the precision for the estimation of a parameter can then be achieved by optimizing over all possible generalized measurement schemes. This optimization defines the quantum Fisher information~\cite{braunstein1994statistical,liu2020quantum}
\begin{equation}
    \mathcal{I}_Q(\Lambda)=\text{Tr}[\rho(\Lambda)L^2(\Lambda)],
    \label{eq:quantumfisher}
\end{equation}
which, in turn, gives us the aforementioned quantum Cramer-R\'ao bound $V(\Lambda)\geq 1/({n \mathcal{I}_Q(\Lambda)})$. Here we have introduced the symmetric logarithmic derivative  $L(\Lambda)$, defined by $\partial_\Lambda\rho(\Lambda)=\{L(\Lambda),\rho(\Lambda)\}/2$.

In the following, we will focus on Gaussian quantum systems~\cite{weedbrook2012gaussian,adesso2005gaussian}. For this particular class of systems, it is convenient to use a phase-space formalism that focuses solely on the first and second moments of the quadratures of the system~\cite{mcmillen2017quantum}. The latter is compactly represented by the covariance matrix $\sigma(\Lambda)$ whose elements are given by $\sigma_{i,j}=\langle \{ r_i,r_j\}\rangle/2-\langle r_i\rangle\langle r_j\rangle$ in terms of the  components of the quadrature operators vector $\hat{\mathbf{r}}$. Moreover, we will also restrict our considerations to local Gaussian measurement on a single Gaussian mode. These are represented by a spectralization of the identity in terms of single-mode Gaussian states characterized by their covariance matrix $\sigma_{\rm m}$. Following~\cite{mcmillen2017quantum}, the classical Fisher information can be written in this case as
\begin{equation}
     \mathcal{I}_C(\Lambda)=\dfrac{1}{2}{\rm{tr}}\left[(\sigma_p^{-1}\partial_{\Lambda}\sigma_p)^2\right],\label{eq:classicalfisherGaussian}
\end{equation}
where $\sigma_p=\sigma(\Lambda)+\sigma_{\rm m}$ is the sum of the covariance matrix of the state of the system $\rho(\Lambda)$ and the one that characterises the measurement POVM. Analogously, the quantum Fisher information for single-mode Gaussian states can be written as~\cite{mcmillen2017quantum} 
\begin{equation}
    \mathcal{I}_Q(\Lambda)=\dfrac{{\det}(\partial_\Lambda\sigma)^2{\rm{tr}}\left[((\partial_\Lambda \sigma)^{-1}\sigma)^2\right]+\frac{1}{2}{\rm{det}}(\partial_\Lambda \sigma)}{2{\rm{det}}\sigma^2-1/8}.\label{eq:quantumfisherGaussian}
\end{equation}

\section{System \& Collapse Mechanism}\label{sec:system}

In the rest of this work, we will investigate the precision limit for parameter estimation in a specific optomechanical set-up affected by a collapse mechanism. 
We consider the two-cavity system {shown in Fig.~\ref{figsetup}.
\begin{figure}[t]
\centering
\includegraphics[width=0.75\linewidth]{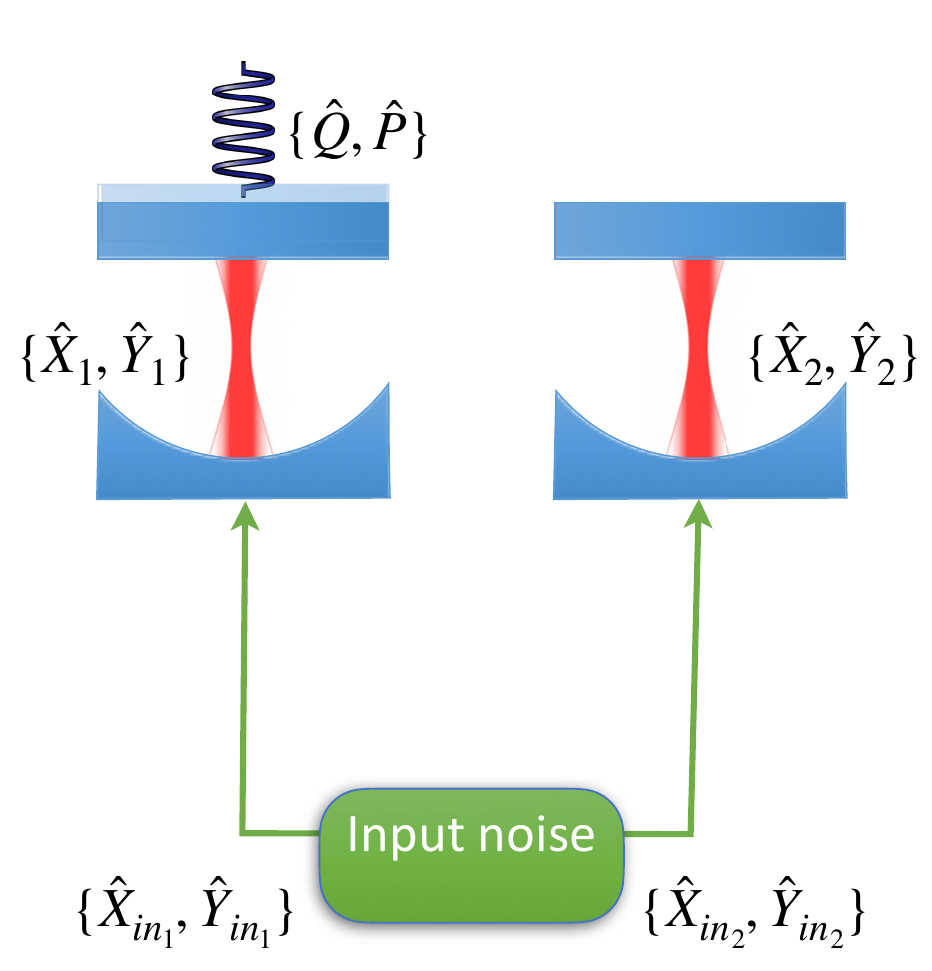}
\caption{Schematic set-up considered in the main text. Two cavities, one of which with a movable end-mirror, are injected with noise described by two Gaussian modes with quadratures $\{\hat{X}_{in_i},\hat{Y}_{in_i}\}$ with $i=\{1,2\}$. The movable end-mirror in cavity 1 represents a Gaussian mechanical mode with quadratures $\{\hat{Q},\hat{P}\}$.}
\label{figsetup}
\end{figure}
This same set-up has been recently analyzed for the purpose to show transient quantum advantages in quantum hypothesis testing for collapse models~\cite{marchese2021optomechanical} and it is inspired by the the quantum reading scheme in~\cite{pirandola2011quantum}. The system consists of an optomechanical cavity and a second, normal cavity used as auxiliary system. The optomechanical cavity is initially pumped with coherent light until it reaches its steady-state. Afterwards, an extra laser beam is injected in both cavities and the output modes are measured either directly or after recombination through a beam-splitter. We study the driven dynamics by comparing classical and quantum sources of input light and local or EPR-like measurements of the output modes. 
We assume that the driving field is strong enough to allow for the linearization of the dynamics. This means that any operators can be split into two parts $\hat{O}=\langle O \rangle+\delta \hat{O}$, where $\langle O \rangle$ is a mean-field part, that behaves classically, and $\delta \hat{O}$ is the quantum fluctuation part. We are going to study the dynamics of the quantum fluctuations~\footnote{For simplicity we will suppress the $\delta$ when indicating the quantum fluctuations' quadrature from here on.}.

In the linear approximation, the Hamiltonian of the system is at most quadratic in the quadratures of the system and the noise we consider has at most linear jump operators. Thus the open system dynamic is Gaussian~\cite{genoni2016conditional}.
In order to use the Gaussian formalism for the parameter estimation framework, we also restrict our analysis to Gausssian measurements. This simply requires addressing the covariance matrix of the CSL-affected optomechanical system $\mathbf{\sigma}(\Lambda)$, whose elements $\sigma_{i,j}=\langle \{ r_i,r_j\}\rangle/2$ are obtained from the zero-mean quantum fluctuations vector $\hat{\mathbf{r}}=\left(\hat{Q},\hat{P},\hat{X}_1,\hat{Y}_1,\hat{X}_2,\hat{Y}_2\right)^{\intercal}$. Here, the first two components $\{\hat{Q},\hat{P}\}$ are the dimensionless quadratures for the mechanical mode of the optomechanical cavity which is modelled as a harmonic oscillator with frequency $\omega_m$ and damping rate $\gamma_m$. The remaining quadratures, $\{\hat{X}_i,\hat{Y}_i\}$, are the optical modes for both cavities $i\in\{1,2\}$.
The time evolution then is given by the Lyapunov-like equation 
\begin{equation}
     \dot{\sigma}= A \sigma +\sigma A^{T} +D,\label{eq:lyapunovequation}
\end{equation}
where $A$ is the drift matrix depending on the physical parameters of the system, and $D$ is the diffusion matrix. The latter accounts for the noises as $  D_{ij}=\dfrac{1}{2}\left[ \langle n_i(t)n_j(t)\rangle+\langle n_j(t)n_i(t)\rangle\right]$, where the quantum noise operators vector is $
\hat{\mathbf{n}}=\left(0, \hat{\xi}+\hat{f}_{\Lambda}, \sqrt{2\kappa}\hat{X}_{in_1},\sqrt{2\kappa}\hat{Y}_{in_1}, \sqrt{2\kappa}\hat{X}_{in_2},\sqrt{2\kappa}\hat{Y}_{in_2}\right)^{\intercal}$, where $\kappa$ is the cavity decay rate, assumed to be the same for both cavity for simplicity~\cite{PhysRevLett.103.213603}. 

Here, we consider the following noise sources:
\begin{enumerate}
\item  The extra input light fields, given by the operators $\{\hat{X}_{in_i},\hat{Y}_{in_i}\}$ for each cavity $i\in\{1,2\}$. 
\item The Brownian noise, described by the noise operator $\hat{\xi}$, characterised by the Markovian correlation functions $\langle\hat{\xi}(t)\hat{\xi}(t')\rangle=2 \dfrac{\gamma_m k_B T}{\hbar \omega_m}\delta(t-t')$. Here $k_B$ is the Boltzmann constant, while $T$ is the temperature  of the surrounding thermal environment.
\item The CSL collapse model described by $\hat{f}_{\Lambda}$ and acting as an extra source of decoherence.
\end{enumerate}
The decoherence due to the collapse mechanism can be effectively ascribed to a stochastic force~\cite{nimmrichter2014optomechanical}, characterized by the two-point correlation function $\langle\hat{f}_\Lambda(t)\hat{f}_\Lambda(t')\rangle=\Lambda\delta(t-t')$. The associated diffusion rate 
\begin{equation}
\Lambda=\dfrac{1}{\hbar \omega_m m}\lambda_{CSL}(\hbar/r_{CSL})^2\alpha
\end{equation} 
depends on the two fundamental CSL parameters, the rate of collapse $\lambda_{CSL}$, and the decoherence length $r_{CSL}$; it involves also a mass-scaling factor $\alpha$ which can be written as
\begin{equation}
    \alpha=\frac{r_c^5}{\pi^{3/2}m_0^2}\int d^3k k_x^2 e^{-r_{c}^2 \mathbf{k}^2}|\tilde{\rho}(\mathbf{k})|^2,
\end{equation}
where $m_0=1$~amu (atomic mass unit) and $\tilde{\rho}(\mathbf{k})=\int d^3 r \rho(\mathbf{r})e^{-i \mathbf{k}\cdot \mathbf{r}}$ is the Fourier transform of the mass density of the system subject to the CSL. 

The observable consequence of the collapse mechanism  is an overheating of the system, mathematically represented by the additional contribution $\hat{f}_{\Lambda}$ to the stochastic Brownian force $\hat{\xi}$. In the diffusion matrix, the collapse diffusion rate $\Lambda$ enters the mechanical mode as an extra thermal constant, added to the Brownian contribution

\begin{equation}
D=\left(\begin{array}{c|c}
\begin{matrix}
 0 & 0 &\\
 0 & 2 \dfrac{\gamma_m k_B T}{\hbar \omega_m}+ \Lambda\\
\end{matrix}

  & \mathbb{O}_{2\times 4} \\ \hline 
\mathbb{O}_{4\times 2}  & \sigma_{in}
\end{array}\right),\label{eq:dmatrix}
\end{equation}
where  ${\sigma}_{in}$ is the 4$\times$4 covariance matrix associated to the driving light input modes $\{\hat{X}_{in_i},\hat{Y}_{in_i}\}$, {and $\mathbb{O}_{n\times m}$ is a $n\times m$ matrix of zeroes}.
Here $\Lambda$ is the unknown parameter at the centre of our parameter estimation effort.

In our set-up, the dynamics of the mechanical system, affected by the collapse mechanism, is indirectly monitored by measuring the cavities' output modes. As already discussed, we restrict the detection of the optical output modes to local Gaussian POVM measurements characterized by the single-mode Gaussian states covariance matrix $\sigma_{\rm m}=R\;{\rm{diag}}(l/2,l^{-1}/2)R^{\rm{T}}$. Here $l\in[0,\infty]$ parametrises the degree of squeezing of the POVM, i.e., $l=\{0,\infty\}$ corresponds to homodyne detection and $l=1$ heterodyne detection, and the matrix $R=\cos(\theta) \mathbbm{1}-i \sin(\theta)\sigma_y$ describes a rotation in phase-space in terms of the Pauli matrix $\sigma_y$ with $\theta$ determining the direction along which the measurement is performed. 
Thus, the total covariance matrix entering the definition of the classical Fisher information~\eqref{eq:classicalfisherGaussian} is given by
\begin{equation}
    \sigma=\sigma(\Lambda) + \sigma_{\rm m}.
\end{equation}
Here, $\sigma(\Lambda)$ is a $2\times 2$ diagonal block of the evolved $6\times 6$ covariance matrix obtained as solution of Eq.~\eqref{eq:lyapunovequation} and it pertains to a single optical cavity mode -- either the one of the first cavity or a linear combination of the two optical modes via a beam splitter as explicated in the following. This will be the only quantity needed to calculate both the classical and the quantum Fisher information. In the latter case, we also just need $\sigma=\sigma(\Lambda)$.

\section{Dynamical analysis}\label{sec:dynamics}

Let us consider the dynamic evolution of the system before it reaches its steady-state.
The initial state of the system is chosen to be the product of the steady-states obtained when only coherent light is pumped into the cavities. Thus, the optomechanical cavity will be in a steady-state of the light field and the mechanical element, while the second cavity will simply be in its ground state. Once this initial state has been reached, it is possible to drive the system by using additional laser light and the output modes of both cavities can be measured. 

We compared two strategies, that we call \textit{classical} and \textit{quantum} according to the choice of input resources and type of measurement performed~\cite{marchese2021optomechanical}.
The \textit{classical} strategy involves two independent thermal input noises as classical sources driving the dynamics. 
This is combined with a local measurement of the optical field of the first cavity $\{\hat{X}_{1},\hat{Y}_{1}\}$. Note that here we refrain from explicitly accounting for the input-output relations needed when one considers the measurement of the output cavity field.
This is a reasonable assumption, given the linearity of the input-output relations, that allows performing measurements on internal the cavity modes without interfering with the output modes~\cite{Paternostro2007}.

The initial covariance matrix for thermal states with mean number of photons $n_1$ and $n_2$ respectively, reads
\begin{equation}
\sigma^{\rm{th}}_{in}=2\kappa\left(\begin{array}{c|c}
(n_1+1/2)\mathbb{I}_{2\times 2}  & \mathbb{O}_{2\times 2} \\ \hline 
\mathbb{O}_{2\times 2}  & (n_2+1/2)\mathbb{I}_{2\times 2}
\end{array}\right).\label{eq:thcovmatrix}
\end{equation}
The local measurement is performed on the first cavity optical mode and thus
 it concerns the $2\times 2$ central diagonal block of the full $6\times 6$ covariance matrix solution of Eq.~\eqref{eq:lyapunovequation}.

The \textit{quantum} strategy, on the other hand, makes use of a two-mode squeezed (TMS) light field as correlated input noise and a quantum measurement of EPR-type quadratures obtained by combining the optical fields of the two cavities with a beam-splitter. 
TMS states are Gaussian states whose covariance matrix, entering Eq.~\eqref{eq:dmatrix}, depends only on the squeezing amplitude $r\geq 0$ and the squeezing angle $\psi_S$ and can be written as
\begin{equation}
\sigma_{in}^{\rm{TMS}}=\kappa\left(
\begin{array}{c|c}
 \cosh 2r \mathbb{I}_{2\times 2}  & \sinh 2r R_{\psi_S}\\
 \hline
  \sinh 2r R_{\psi_S}& \cosh 2r \mathbb{I}_{2\times 2}
\end{array}
\right),\label{eq:TMScovmatrix}
\end{equation}
where 
\begin{equation}
    R_{\psi_S}=\begin{pmatrix}
     \cos\psi_S & \sin\psi_S\\
     \sin\psi_S & -\cos\psi_S
    \end{pmatrix}.
\end{equation}

The EPR-like measurements correspond to measuring a linear combination of the optical modes of the cavities obtained via a 50:50 beam-splitter giving rise to modes with quadratures
\begin{equation}
\hat{q}_{\mp}=\dfrac{\hat{X}_{1}\mp\hat{X}_{2}}{\sqrt{2}},\quad\hat{p}_{\pm}=\dfrac{\hat{Y}_{1}\pm\hat{Y}_{2}}{\sqrt{2}}.
\end{equation}

In terms of covariance matrix elements, it means that the $4\times 4$ submatrix $\sigma_{1,2}(t)$ of the solution to Eq.~\eqref{eq:lyapunovequation}, representing the covariance matrix of the two optical cavity modes at a generic time $t$, has to go through a simplectic transformation describing the modes recombination via the beam-splitter~\cite{genoni2016conditional}
\begin{equation}
    \sigma^{\rm{EPR}}=\hat{S}\sigma_{1,2}(t)\hat{S}^{\rm{T}}.\label{eq:EPRcovmatrix}
\end{equation}
The operator $\hat{S}=e^{\Omega \hat{H}_{BS}}$ satisfies the equation $\hat{S}\Omega\hat{S}^{\rm{T}}=\Omega$, where $\Omega=\bigoplus^2_{j=1}\begin{pmatrix}
 0 & 1\\
 -1 & 0\\
\end{pmatrix}$ is the symplectic matrix and $H_{BS}=\frac{\varphi_{BS}}{2}(\hat{a}^\dagger\hat{b}-\hat{a}\hat{b}^\dagger)$ the beam-splitter Hamiltonian with $\varphi_{BS}$ the beam-splitter angle. Here, $\{\hat{a}^\dagger,\hat{a}\}$ ($\{\hat{b}^\dagger,\hat{b}\})$ are the creation and annihilation operators for the two optical cavity modes respectively~\cite{ferraro2005gaussian}.

These two measurement strategies are analogous to the ones used in the quantum hypothesis testing employing the same optomechanical set-up in~\cite{marchese2021optomechanical}. In this case, however, since we are interested in the precision limit to the estimation of the CLS parameter, we look at the (classical) Fisher information for the two strategies that we have discussed. The classical Fisher information is obtained from Eq.~\eqref{eq:classicalfisherGaussian}. While $\sigma_{\rm m}$ characterises the measurement on the single optical mode, $\sigma(\Lambda)$ is the CSL-affected covariance matrix obtained by (i) solving Eq.~\eqref{eq:lyapunovequation} with either the input noise from Eq.~\eqref{eq:thcovmatrix}, for the classical strategy, or Eq.~\eqref{eq:TMScovmatrix}, for the quantum strategy; and (ii) either focusing on the $2\times 2$ submatrix corresponding to the optical mode of the first cavity, for the classical strategy, or one the $2\times 2$ covariance matrix of one of the optical modes emerging from the beam-splitter mixing the optical modes of the two cavities, for the quantum strategy. \\
In Fig.~\ref{fig:QvsC} we show the classical Fisher information for the two strategies in function of time. We observe that for early times, the \textit{quantum} scheme gives a higher value of the Fisher information than the \textit{classical} one. This translates in a lower bound, with respect to the classical scheme, on the precision of the estimation of the CSL parameter $\Lambda$.  
The precision to which we can estimate the parameter $\Lambda$ using classical input states and measurements can be overcome at short times by using non-classical resources, namely TMS states and EPR measurements. We also observe that the quantum advantage is lost at later times. This is expected for systems subjected to decoherence arising from the thermal noises~\cite{mirkhalaf2022operational}.
These results are in agreement with those obtained in~\cite{marchese2021optomechanical} where, considering the same set-up with the same choice of parameters, a \textit{quantum} advantage at short times was proven for (quantum) hypothesis testing aimed at certifying the presence of the CLS collapse mechanism.

\begin{figure}
 \centering
 \begin{minipage}[c]{\columnwidth}
    \centering
    \includegraphics[width=3.0in]{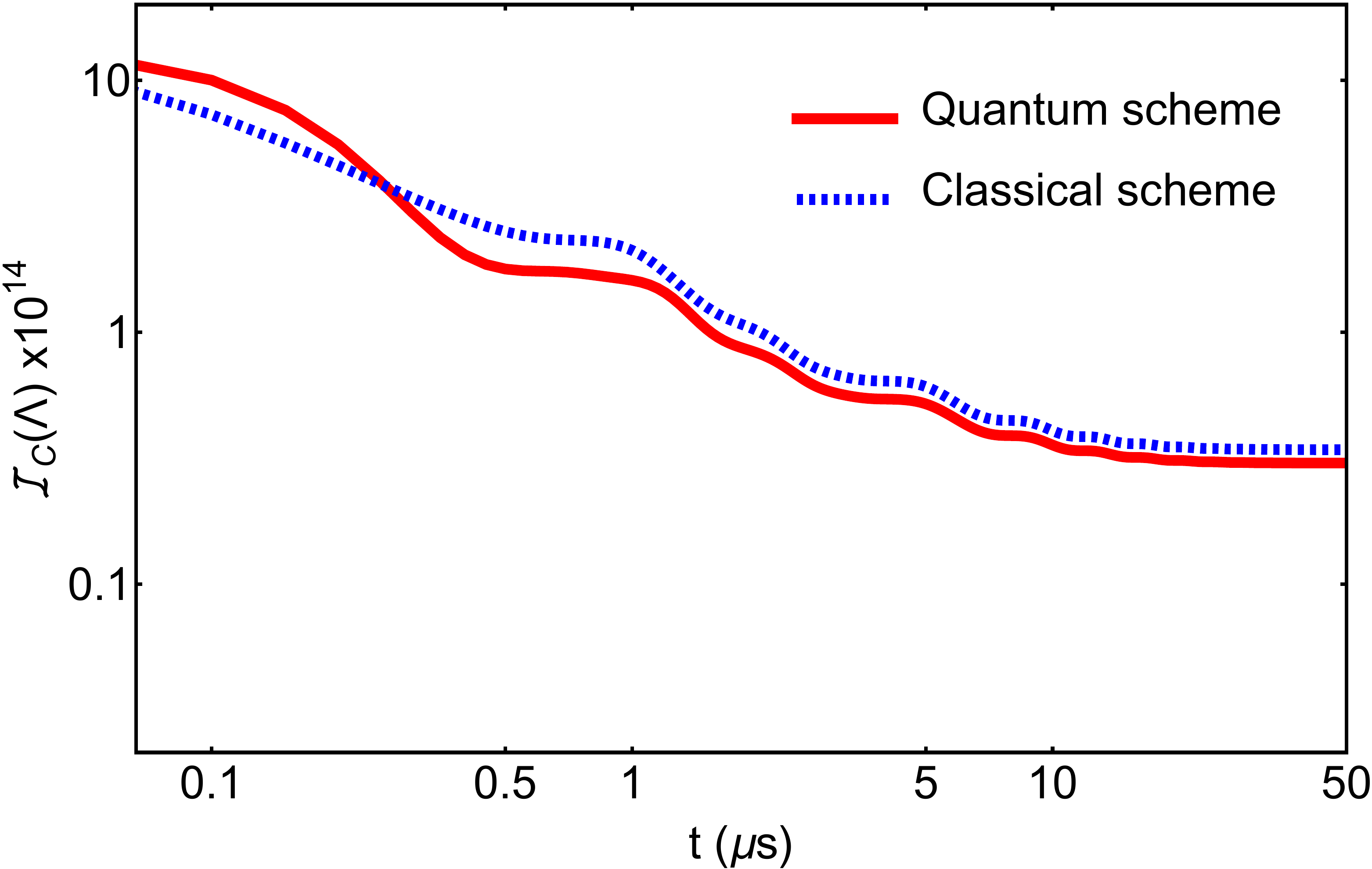}
     \includegraphics[width=3.0in]{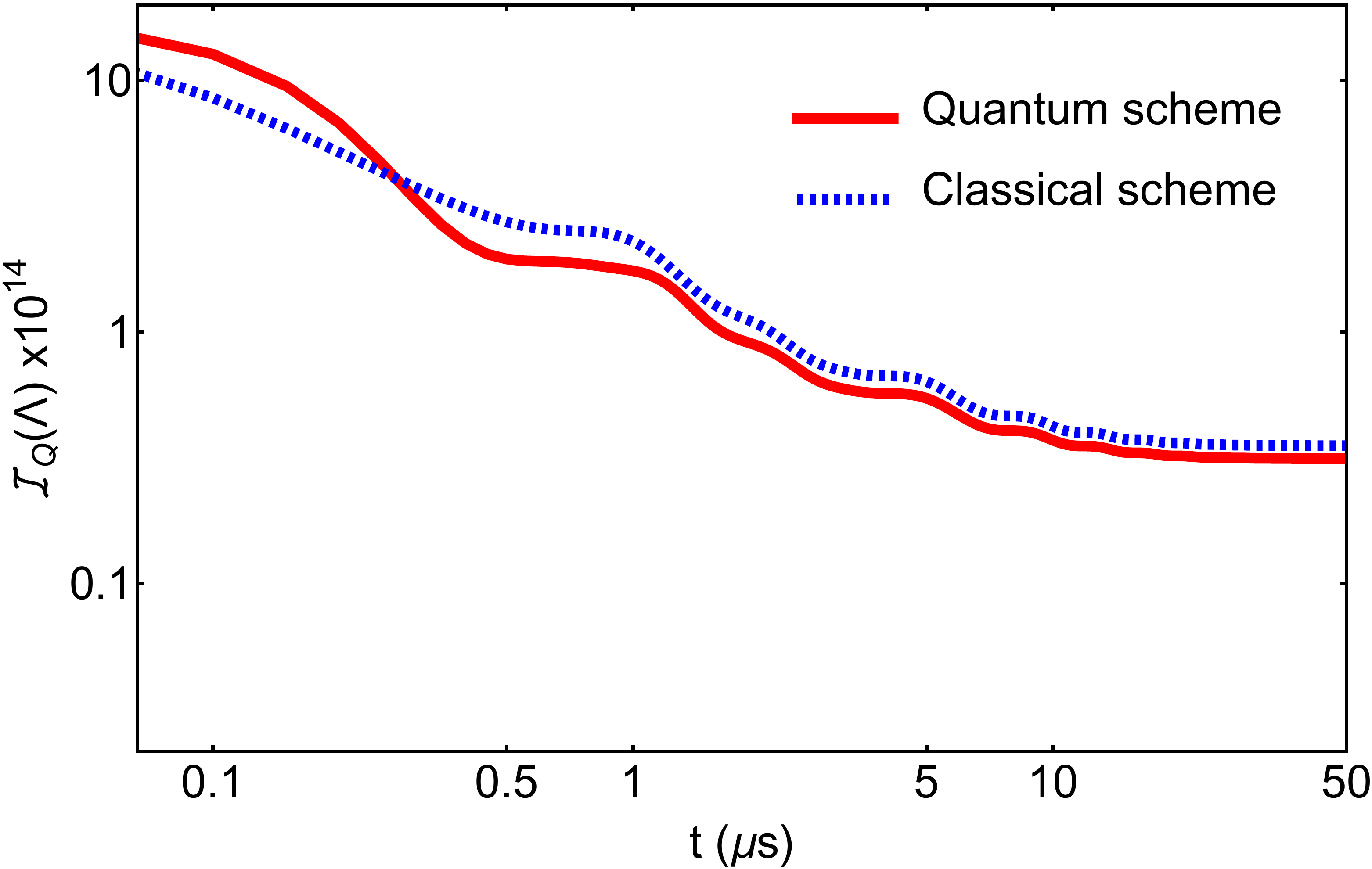}
    \caption{
    Classical and Quantum Fisher information for the \textit{quantum} and the \textit{classical} schemes. For the \textit{quantum} scheme, we set the squeezing angle $\psi_S=\pi$ for the input TMS light, and $\varphi_{BS}=\pi/4$ for the beam-splitter angle employed in the EPR measurement. We used $\Lambda=10^{6}$, which results from assuming $r_{CSL}=100\,{\rm{nm}}$ and Adler's collapse rate~\cite{adler2007lower} $\lambda_{CSL}\equiv \lambda_A=10^{-9}{\rm{s^{-1}}}$. The parameters for the measurement covariant matrix $\sigma_{\rm m}$ are set to $l=1$ and $\theta=0$. Only at small times, up to $t\sim 0.25\rm{\mu s}$, the \textit{quantum} scheme brings an advantage over the \textit{classical} one.}
    \label{fig:QvsC}
 \end{minipage}
 \end{figure}

\section{Steady-state analysis}\label{sec:steadystate}

Having considered the transient dynamics, in this Section we perform a steady-state analysis. In line with the previous discussion, we observe that the best performance is always obtained with a classical scheme.

Once the full system reaches a steady-state, all memory about the initial state is lost. However, according to the kind of input noises we subjected the system to -- either thermal or TMS light--, the dynamics will drive the systems to different steady-states. We compute both the classical and the quantum Fisher information at the steady-state using the covariance matrix $\sigma_{\rm ss}$ obtained as the solution of Eq.~\eqref{eq:lyapunovequation} when setting the right-hand side to zero, i.e., 
\begin{equation}
     A \sigma_{\rm ss} +\sigma_{\rm ss} A^{T} =-D.\label{eq:lyapunovequationsteady}
\end{equation}

\begin{figure}[t]
    \centering
    \includegraphics[width=0.45\textwidth]{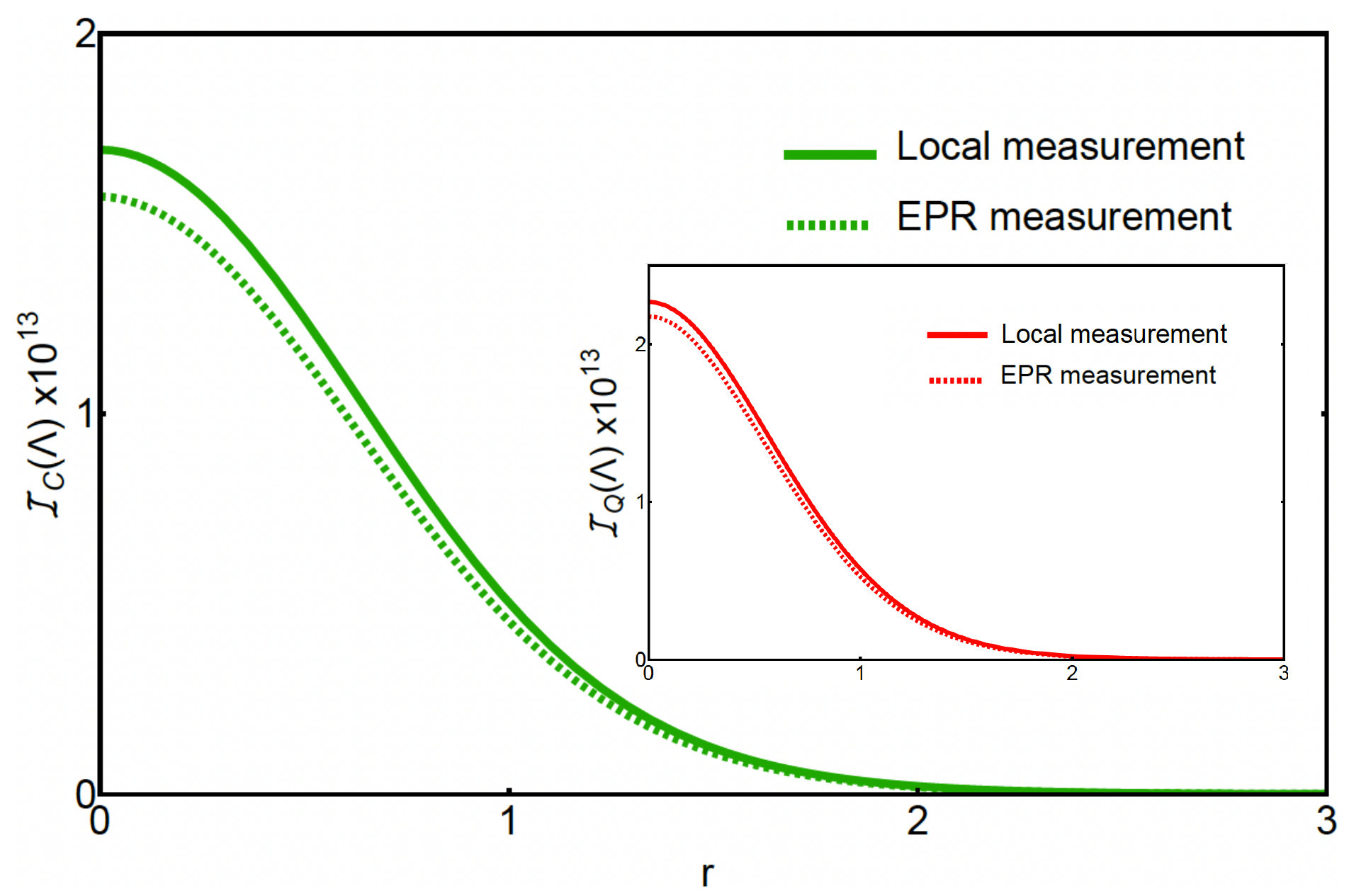}
    \caption{Classical Fisher information at the steady-state against the squeezing parameter $r$ of the TMS input light. 
    We compare two measurement scheme: local measurements (blue curve) and EPR measurements (green curve). The squeezing angle of the input TMS state is set to be $\psi_S=\pi$. However, for the local measurements this does not change the Fisher information. The EPR measurement scheme uses a beam-splitter angle $\varphi_{BS}=\pi/4$ to combine the two otical cavity modes. In both cases the classical Fisher information vanishes with increasing squeezing parameter $r$. The inset shows the same plots for the quantum Fisher information which, qualitatively, gives the same results.}
    \label{fig:CfisherlocalvsEPR}
\end{figure}
In Fig.~\ref{fig:CfisherlocalvsEPR} we show the classical Fisher information at the steady-state reached with TMS-input-noise driven dynamics. The plots are in function of the squeezing parameter $r$ of the input light field. We compare the two measurement schemes, the \textit{quantum} one, i.e., EPR measurement, and the \textit{classical} one, that employs local measurements. Higher values of the Fisher information are obtained for lower values of the squeezing parameter. In particular, the maximum is obtained when $r=0$ and for local measurements. In other words, when there is no 2-mode squeezing in the input-noise and we only focus on the first cavity we obtain the minimum error in the estimation of the CSL parameter $\Lambda$. This corresponds to the case in which we drive the cavities with just coherent light, which can be considered a classical input light field and, as a matter of fact, we completely neglect the second optical cavity. In particular, we see that neither input-noise 2-mode squeezing nor EPR-like measurements of the optical cavity modes can lead to an advantage in the estimation of the CSL parameter. 
Therefore, we conclude that the use of quantum measurements and input noise is not helpful in the estimation of the CSL parameter $\Lambda$ at the steady-state where instead local measurements and vacuum input noise lead to the best estimate.

\section{Conclusions}\label{sec:conclusions}

In this work, we re-considered a previously proposed optomechanical set-up, showing an advantage for quantum hypothesis testing directed at investigating collapse model dynamics, from the point of view of parameter estimation.
By investigating the non-equilibrium dynamics of the system, we find that a combination of quantum correlated input-noises and EPR-like measurements provides an advantage in the estimation of the CLS parameter $\Lambda$ at short times compared to a classical strategy. This corroborates the result previously obtained for the hypothesis testing protocol~\cite{marchese2021optomechanical}. Nonetheless, this advantage is lost at the steady-state. Indeed, at the steady-state a classical measurement scheme and an uncorrelated vacuum input-noise outperform EPR-like measurement and quantum correlated 2-mode squeezed input-noises. This is valuable information for any experimental effort aimed at nailing down the potential occurrence of collapse-like mechanisms on the dynamics of a quantum system. In particular, it highlights the benefits that a non-equilibrium regime provides in magnifying the advantages provided by quantum resources. 

\section*{Acknowledgements}
MMM is grateful to the EPSRC Large Baseline Quantum-Enhanced Imaging Networks (Grant No. EP/V021303/1), and the EPSRC Quantum
Communications Hub (Grant No. EP/M013472/1). AB and MP acknowledge support from the Horizon Europe EIC Pathfinder project  QuCoM (Grant Agreement No.\,101046973). A.B. is supported by the Deutsche Forschungsgemeinschaft (DFG, German Research Foundation) project number BR 5221/4-1. MP thanks the European Union's Horizon 2020 FET-Open project  TEQ (Grant Agreement No.\,766900), the Leverhulme Trust Research Project Grant UltraQuTe (grant RPG-2018-266), the Royal Society Wolfson Fellowship (RSWF/R3/183013), the UK EPSRC (EP/T028424/1), and the Department for the Economy Northern Ireland under the US-Ireland R\&D Partnership Programme (USI 175 and USI 194). 

\bibliographystyle{ieeetr}

\bibliography{references2}

\end{document}